\documentclass[final,5p,times,twocolumn,authoryear]{elsarticle}

\usepackage{amssymb}
\usepackage{amsmath}
\usepackage{booktabs}    %
\usepackage{multirow}  
\usepackage[utf8]{inputenc} %
\usepackage{circledsteps}

\usepackage{caption} 
\usepackage[normalem]{ulem}
\usepackage{enumitem}

\usepackage[hidelinks]{hyperref}
\usepackage{listings}
\usepackage{xcolor}
\newcommand*{\system}{ZoomTable}
\newcommand{\cqy}[1]{\textcolor{black}{#1}}

\journal{Visual Informatics}

\begin{document}

\begin{frontmatter}

\title{ZoomTable: Interactive Exploration of Data Facts in Hierarchical Tables \\ via Semantic Zooming} %

\author[label1]{Qiyang Chen}
\author[label1]{Guozheng Li}
\author[label1]{Xingqi Wang}
\author[label1]{Gerile Aodeng}
\author[label2]{Min Lu}
\author[label1]{Chi Harold Liu}

\affiliation[label1]{organization={School of Computer Science, Beijing Institute of Technology}, addressline={No. 5 Zhongguancun South Street}, city={Beijing}, postcode={100081}, country={China}}
\affiliation[label2]{organization={School of Architecture and Urban Planning, Shenzhen University}, addressline={No. 3688 Nanhai Avenue}, city={Shenzhen, Guangdong Province}, postcode={518060‌}, country={China}}

\begin{abstract}
Hierarchical tables are an important structure for organizing data with inherent hierarchical relationships.
Existing studies have extensively explored methods for data fact exploration from tabular data. In particular, some studies have directly integrated visual data facts into the original table structure to support in-situ exploration, because embedding data facts within the table context can reduce cognitive load by minimizing attention shifts. 
However, embedding a large amount of extracted data facts into the limited space of hierarchical tables often leads to layout conflicts, hindering effective exploration.
To address this issue, we propose an interactive exploration paradigm for hierarchical table data facts based on semantic zooming and develop an interactive visualization system, ZoomTable. 
The ZoomTable system employs semantic zooming as the interaction method, combined with a data-fact layout method and a data fact recommendation mechanism. 
This combination not only resolves layout conflicts, but also supports users in coherently exploring multidimensional data facts at different scales. 
A case study and a user experiment further validate the practicality and efficiency of ZoomTable in real-world data fact exploration scenarios.
\end{abstract}


\end{frontmatter}

\section{Introduction}
In the domain of data analysis and visualization, hierarchical tables serve as an important data representation technique widely adopted to illustrate complex data~\citep{e5_zhang_acl24,AIT-QA,2022-hitab,cqy_rag_2025}.
Hierarchical tables organize data through multilevel row and column headers and typically contain abundant data facts~\citep{CoInsight2024Li}. 
Visual exploration of data facts aims to identify, analyze, and organize such facts to help users acquire key insights~\citep{Voyager1,CoInsight2024Li,Groot_VIS2024,Voyager2} efficiently.
Therefore, designing effective methods to support the in-depth exploration of these facts in the context of hierarchical tables is an important research problem.

Existing studies have investigated approaches for visual exploration of data facts from tabular data, which can be broadly divided into two groups.
The first group automatically extracts data facts using algorithmic techniques and generates structured visual summaries, providing a basis for user understanding and further exploration~\citep{DataShot,LuJunhua_Scrollytelling,ChartStory_2023}.
The second group supports data fact exploration through interactive methods, such as natural language interaction~\citep{LEVA_csm_TVCG25,LIDA2023} or novel visualizations that enable interactive exploration~\citep{CoInsight2024Li,DataShot}.
However, existing work generally overlooks mapping visual elements back to their originating tabular regions. 
Previous studies~\citep{The_table_lens,HiTailor,Taggle} have demonstrated that embedding visualizations in the original table context effectively reduces cognitive load, thus enhancing user comprehension and operational efficiency. 
Consequently, when exploring data facts in hierarchical tables, we seek a method that tightly integrates visual representations with the original table context.

However, conducting efficient data fact exploration within the original table context remains a major challenge for hierarchical tables, primarily because such tables contain a large number of data facts with complex interrelations. Directly embedding data facts into tables can lead to information overload and layout conflicts, thus affecting the integrity of the data presentation and reducing the quality of the interactive experience. Existing static embedding approaches often fail to effectively organize and present data facts in such contexts.

\begin{figure*}[!t]
    \centering
    \includegraphics[width=1\linewidth]{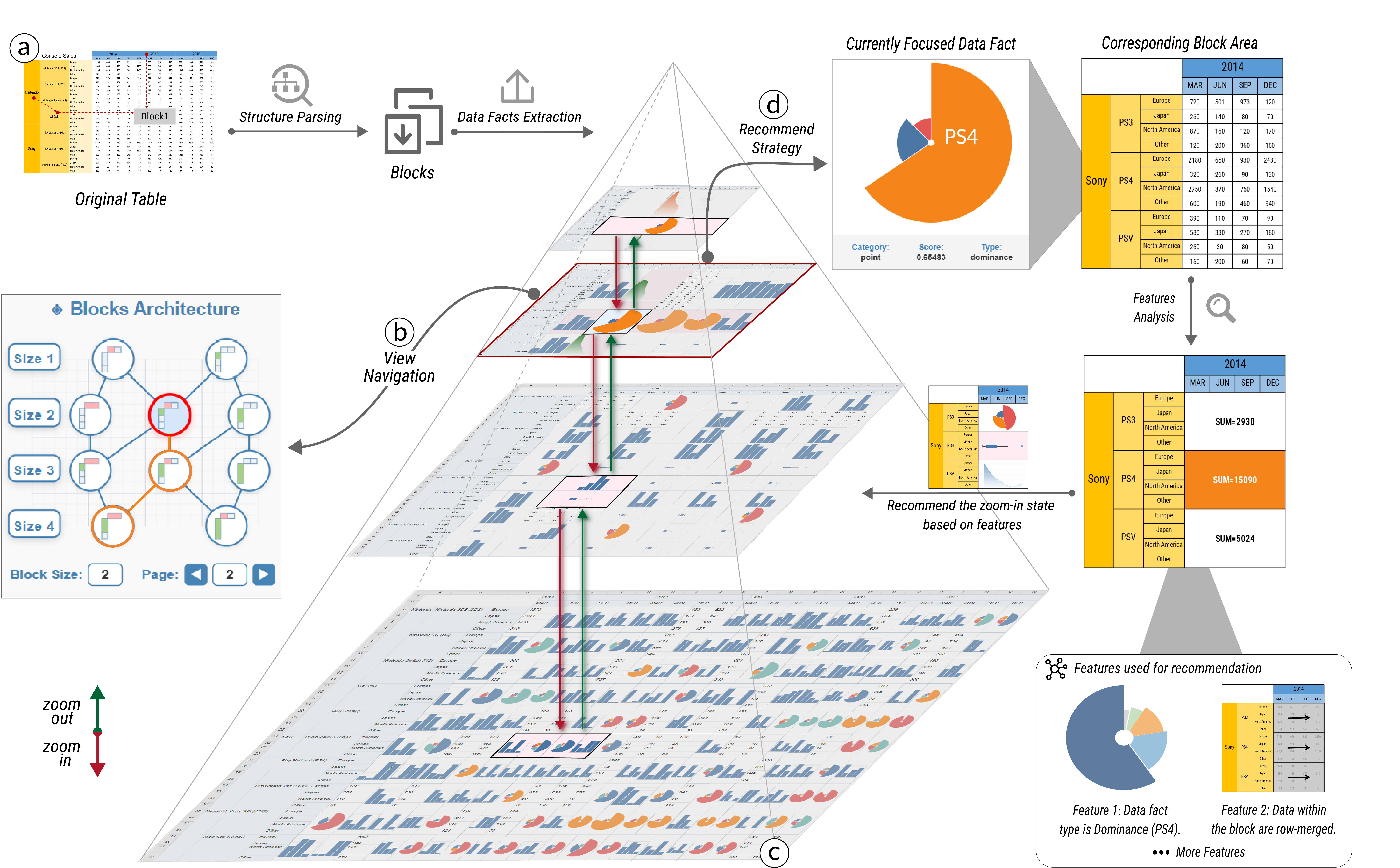}
    \caption{The pipeline of the interactive data facts exploration paradigm for hierarchical tables based on semantic zooming. (a) Parsing the original hierarchical tables and extracting data facts; (b) the navigation panel supporting users in understanding the characteristics of the current state and the exploration path; (c) the most detailed state based on the semantic zooming interaction, where zoom-in and zoom-out operations enable exploration of data facts at different scales; (d) the recommendation strategy, which recommends the data facts to be displayed after zooming based on the characteristics of the currently focused data facts.
}
    \label{fig:pipeline}
\end{figure*}

To address the challenge and support data fact exploration in hierarchical tables, we propose an interactive exploration paradigm (Fig.~\ref{fig:pipeline}) based on semantic zooming interaction consisting of three steps:  
(1) Parsing Table Structure: parsing the hierarchical structure of table headers, locating data blocks, and automatically extracting multi-scale data facts;  
(2) Laying Out Data Facts: mapping data facts to multiple distinct states according to header characteristics, thereby alleviating layout conflicts within the limited table space;  
(3) Exploration based on Semantic Zooming: supporting cross-scale navigation within the original table context and, with the aid of a state recommendation mechanism, suggesting appropriate state at the zoomed level based on the currently focused data facts to improve the efficiency of exploration path construction.  
Guided by the design considerations proposed by domain experts, we develop the visualization system \textbf{\system{}}, which implements these steps as an integrated interactive workflow.

To validate the usability and effectiveness of ZoomTable, we conducted a two-stage evaluation consisting of a case study and a user experiment. 
First, in the case study, we collaborated with an experienced data analyst to explore hierarchical tables on business sales from a real world scenario. 
The results show that ZoomTable supports insightful data fact exploration within the original table context and demonstrates practical applicability. 
Second, we conducted a user experiment that included both comparative and ablation studies. 
In the comparative experiment, we compared ZoomTable with vizGPT\footnote{https://vizgpt.ai/} and CoInsight~\citep{CoInsight2024Li}, asking participants to construct exploration paths under assigned tasks and evaluating outcomes using both quantitative metrics and subjective ratings; the results indicate that ZoomTable outperforms the baselines in both exploration path efficiency and user experience. 
In the ablation experiment, we individually removed the state recommendation mechanism, fact type filtering, and layout-structure visualization for comparison, and the results show that all three modules play critical roles in improving exploration efficiency. In general, ZoomTable exceeds existing methods in terms of exploration efficiency, interface design, and task experience.
In summary, the primary contributions of this paper are as follows.

\begin{itemize}
\item An interactive exploration paradigm for hierarchical tables, leveraging semantic zooming combined with data fact layout to support cross-scale exploration;

\item The \system{} system based on the interactive paradigm and design considerations derived from domain experts;

\item A case study and a user experiment to validate the effectiveness of \system{}, showing that it outperforms existing methods in exploration efficiency and user experience.
\end{itemize}

\section{Related Work}
This section presents an overview of relevant research from two aspects: tabular data visualization and data facts exploration.
This section presents an overview of relevant research from two aspects: tabular data visualization and data facts exploration.

\subsection{Tabular Data Visualization}
Data visualization for tables is the design of visual elements derived from the table’s characteristics, aiding users’ comprehension and analysis.
Existing methods largely fall into two approaches: non-embedded and embedded. 

\textbf{Non-embedded.} In non-embedded methods, visualizations are separated from the table structure. TaCo~\citep{TACO} uses diverse visuals to show structural and content differences across table versions produced by transformations, but it is not tailored to specific data characteristics. By contrast, DataShot~\citep{DataShot} focuses on table-data characteristics and automates transforming the original table into vividly designed Fact sheets; Calliope~\citep{Calliope} likewise automates visual data-story construction for spreadsheets. Although efficient and visually appealing, these methods provide limited support for users’ subjective exploration of the original table. Several works add interaction within this non-embedded paradigm: \cqy{Wiens et al. group data into information sets so users can choose suitable visualizations~\citep{Vitalis};} Microsoft Power BI~\citep{Microsoft_Power_BI} and ADVISor~\citep{ADVISor} generate interactive visualizations from natural-language input; and AdaVis~\citep{AdaVis} and Hao et al. extend the scope to tabular datasets and tabular databases~\citep{HaoIALP2024}, respectively. Yet, because visuals remain separate from the table structure, these approaches still struggle to link visuals intuitively to the data and its structural positions.

\textbf{Embedded.} To strengthen this link, embedded methods map visual elements directly onto the original table. \cqy{Perin et al.} encode cell values using matrix-analysis techniques to improve visual clarity~\citep{Charles2014}. Taggle~\citep{Taggle} dynamically aggregates table subsets and visually encodes each row. While Taggle scales to large tables, it struggles with certain formats, especially hierarchical tables with nested headers. For such structures, HiTailor~\citep{HiTailor} parses the table via an abstract model and supports embedding charts in user-selected regions, but limited table space makes overlapping visualizations a time-consuming process. InsigHTable~\citep{InsigHTable} proposes a deep reinforcement learning framework to recommend visualizations in hierarchical tables, enabling user selection based on model suggestions, yet outputs can be unstable and inefficient and do not resolve spatial constraints when embedding visualizations.

\subsection{Data Facts Exploration}
Visual data‐fact exploration focuses on discovering and presenting a series of data facts, that is, statistically significant patterns or features in the data~\citep{Foresight}, and linking them meaningfully~\citep{definition_vis_story}. Tang et al. study the automatic extraction of data facts from multidimensional data~\citep{TangBo2017}, overcoming the limitations of traditional OLAP tools that require manual selection of grouping attributes and query targets. VisPilot~\citep{VisPilot} further accounts for potential drill-down fallacies during data‐fact exploration and proposes a more efficient solution. Ma et al. design a new scoring function for data facts to ensure high‐quality extraction results~\citep{MaPingchuan_MetaInsight}.

With the definition and extraction of data facts well established, subsequent research investigates how these facts can drive the exploration process. For example, \cqy{Ehsan et al. propose three view recommendation schemes that help analysts quickly locate views relevant to their current task among a large set of candidates~\citep{Humaira}.} SpotLight~\citep{Camille} ranks annotated visualizations by relevance and groups them into data-fact clusters to highlight inherent connections. In addition, Lu et al. enumerate data facts in a dataset and organize them in an ordered manner~\citep{LuJunhua_Scrollytelling}. CoInsight~\citep{CoInsight2024Li} focuses on hierarchical tables, extracts data facts, and constructs association graphs to link these facts, thereby helping analysts deriving insights from complex tabular data.

These studies advance the automation and intelligence of data fact extraction, recommendation, and visualization from different perspectives. However, organizing and exploring data facts in hierarchical tables still poses challenges. Although CoInsight~\citep{CoInsight2024Li} leverages header structures to build associations among data facts, it lacks a direct mapping between the facts and the original table context, which may add cognitive load for users during analysis.

\section{Exploration Paradigm} 
\label{sec:Method}
In this section, we elaborate on the proposed interactive exploration paradigm for hierarchical table data facts. This paradigm is implemented through three steps: \textbf{Parsing Hierarchical Table Structure}, \textbf{Laying Out Data Facts}, and \textbf{Applying Semantic Zooming}. Each step provides the necessary support and foundation for the next step, enabling the exploration and analysis of data facts within the original hierarchical table context.

First, \textbf{parsing hierarchical table structure} serves as the foundational step, where the nested structure of the table headers is parsed to accurately identify each data region (table block) within the table. This process provides positioning support for the subsequent extraction of data facts.
Next, \textbf{laying out data facts} builds on the previous structure parsing by extracting data facts with various dimensions and statistical characteristics. These data facts are then systematically embedded into the original hierarchical table space according to a designed layout architecture. This layout design aims to address layout conflicts between data facts and provides structural support for subsequent interactive exploration.
Finally, \textbf{applying semantic zooming}, as an interaction support mechanism, allows users to flexibly explore data facts at different zoom levels within the original table structure. Meanwhile, the system combines a state recommendation mechanism based on the currently focused data, recommending the zoomed states and the corresponding data fact charts within those states.

Through these three steps, the proposed paradigm supports multi-scale exploration of data facts in hierarchical tables. The workflow of this paradigm is shown in Fig.~\ref{fig:pipeline}, and the following describe the detailed implementation of each step, in line with the full process.

\subsection{Parsing Hierarchical Table Structure} \label{sec:Hierarchical-Table-Structure-Parsing}

\begin{figure}
    \centering
    \includegraphics[width=1.0
    \linewidth]{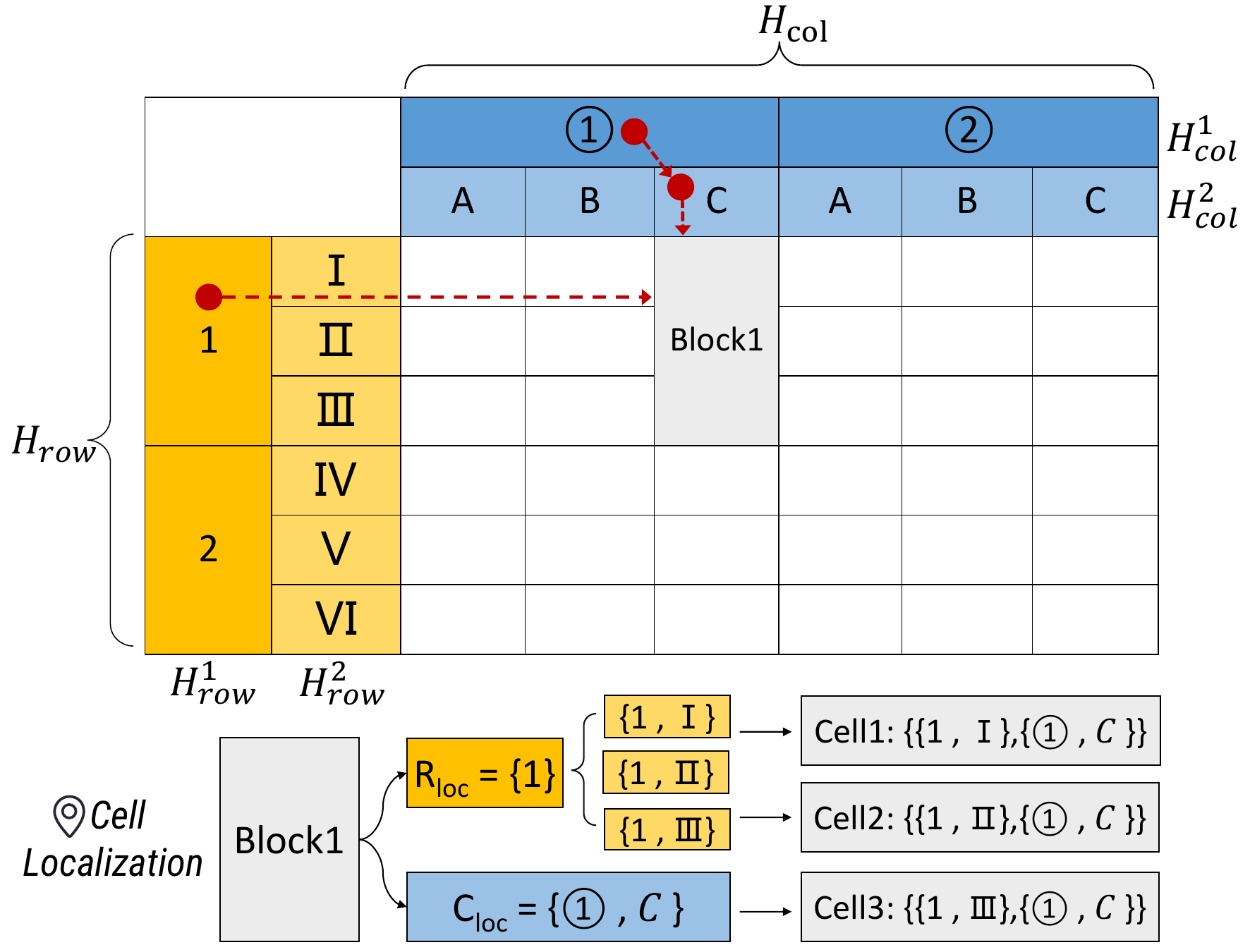}
    \caption{This figure illustrates the method for parsing the structure of hierarchical tables. Specifically, \( H_{\text{row}} \) and \( H_{\text{col}} \) represent the row and column hierarchical structures, respectively.
}
    \label{HTable_Structure}
    \vspace{-1em}
\end{figure}

Hierarchical tables differ from flat tables due to their nested header structures. 
In flat tables, cells can be directly located using row and column indices, 
whereas in hierarchical tables, identifying cells or regions requires more fine-grained header information. 
This section elaborates on the methodology for analyzing hierarchical tables and representing table regions (see Fig.~\ref{fig:pipeline}(a)).

We define a hierarchical table \( T_h \) as a tuple consisting of three elements \cqy{\( \{ H_{\text{row}}, H_{\text{col}}, D \} \). \( H_{\text{row}} \) represents the hierarchical structure of the row header in \( T_h \), \( H_{\text{col}} \) represents the hierarchical structure of the column header in \( T_h \), and \( D \) represents the data entries in the body of the table.
Subsequently, \( H_{\text{row}} \) can be refined to \( \{ H_{\text{row}}^1, H_{\text{row}}^2, \dots, H_{\text{row}}^r \} \), where \( r \) represents the number of levels in the row header of \( T_h \), and \( H_{\text{row}}^r \) refers to the \( r \)-th layer of the row header hierarchical structure. 
Similarly, \( H_{\text{col}} \) is \( \{ H_{\text{col}}^1, H_{\text{col}}^2, \dots, H_{\text{col}}^s \} \), where \( s \) represents the number of levels in the column header of \( T_h \), and \( H_{\text{col}}^s \) represents the \( s \)-th layer of the column header hierarchical structure.} This multi-layered header structure allows hierarchical tables to organize and display complex data in a more refined and flexible manner, meeting diverse data processing and analysis requirements.

As shown in Fig.~\ref{HTable_Structure}, our paradigm parses the header structure of hierarchical tables based on the above definition. Furthermore, a region in the table body that can be described by row and column header structures is referred to as a block. To describe block location precisely, a binary tuple \( B_{\text{loc}} = \{ R_{\text{loc}}, C_{\text{loc}} \} \) is introduced as a location identifier. Here, \( R_{\text{loc}} = \{ R_1, R_2, \dots, R_m \} \) is used to describe the location of the row header, where \( m \) is the deepest level of the row header hierarchy corresponding to this block, and \( R_m \) represents the node label at the \( m \)-th level of the row header. \( C_{\text{loc}} = \{ C_1, C_2, \dots, C_n \} \) is used to describe the location of the column header, where \( n \) is the deepest level of the column header hierarchy corresponding to this block, and \( C_n \) represents the node label at the \( n \)-th level of the column header.
For example, the binary location tuple for Block1 is \( R_{\text{loc}} = \{ 1 \} \) and \( C_{\text{loc}} = \{ \Circled{1}, C \} \), i.e., Block1 = \( \{ \{ 1 \}, \{ \Circled{1}, C \} \} \).
This method of block location using a binary tuple can accurately determine the location of specific data regions in hierarchical tables and support the processing and analysis of complex hierarchical tabular data.

\subsection{Laying Out Data Facts}
This step aims to extract data facts from different blocks of raw data and present them to users in an intuitive layout. Therefore, it is crucial to effectively extracting data facts and organizing them in the table context. We divide this method into two parts: \textbf{data fact extraction} and \textbf{layout architecture design}.

\subsubsection{Data Fact Extraction} \label{sec:insight-extraction} 
The parsing method enables to identify data blocks based on the hierarchy of row and column headers. 
This section highlights the process of extracting underlying data facts from these blocks (see Fig.~\ref{fig:pipeline}(a)).

In the context of a hierarchical table \( T_h \), prior to extracting data facts, the specific content of each block is first retrieved. 
For each block, the row and column header hierarchies are traversed to obtain location information \( B_{\text{loc}} \). 
Subsequently, based on \( B_{\text{loc}} \), the corresponding data cells within the block are located and extracted.
When processing a block, \( B_{\text{loc}} \) is decomposed into the location of the row header \( R_{\text{loc}} \) and the column header \( C_{\text{loc}} \). 
Both \( R_{\text{loc}} \) and \( C_{\text{loc}} \) are recursively traversed to derive complete header paths down to the deepest level.
For both \( R_{\text{loc}} \) and \( C_{\text{loc}} \), recursion starts from the current deepest level and proceeds down the row and column hierarchies of \( T_h \).
At each recursive step, all child nodes at the current level are enumerated and recursion continues until the deepest header level of \( T_h \) is reached. This process allows us to obtain the complete recursive header paths.

Assume that the row and column headers of \( T_h \) have \( r \) and \( s \) levels, respectively.
For the block, its \( R_{\text{loc}} = \{R_1, R_2, \dots, R_m\} \) and \( C_{\text{loc}} = \{C_1, C_2, \dots, C_n\} \). Specifically, the recursive path of the row header \( P_l \) is defined as:

\begin{equation}
P_l = (R_1, R_2, \dots, R_m, h_{l_{m+1}}, \dots, h_{l_r}),
\end{equation}
where \( h_{l_r} \) represents the choice at the \( r \)-th level of the row header node, with the sequence \( \{R_1, R_2, \dots, R_m\} \) remaining fixed throughout the traversal process. Similarly, the recursive path of the column header \( P_t \) is defined as:

\begin{equation}
P_t = (C_1, C_2, \dots, C_n, h_{t_{n+1}}, \dots, h_{t_s}),
\end{equation}
where \( h_{t_s} \) denotes the choice at the \( s \)-th level of the column header node, with the sequence \( \{C_1, C_2, \dots, C_n\} \) remaining fixed.

By combining the row-header path \( P_l \) and the column-header path \( P_t \), a cell within the block can be uniquely identified. A cell \( C \) can be represented at the intersection of the row header path \( P_l \) and the column header path \( P_t \) as \( C = (P_l, P_t) \). This ensures that each cell can be uniquely located through its recursive row and column paths. Consequently, for each combination of the recursive path in rows and columns \( P_l \) and \( P_t \), we can accurately locate a specific table cell by recursion, thus extracting and analyzing the content of each block.

Before extracting data facts, we transform cell information according to block structures, such as \textbf{Merging} (row/column aggregation, e.g., sum, extrema, mean) and \textbf{Extraction} (retrieving homogeneous row/column sets of equal size). These transformations enable better organization of the block’s content and structure.

\begin{figure}
    \centering
    \includegraphics[width=0.9\linewidth]{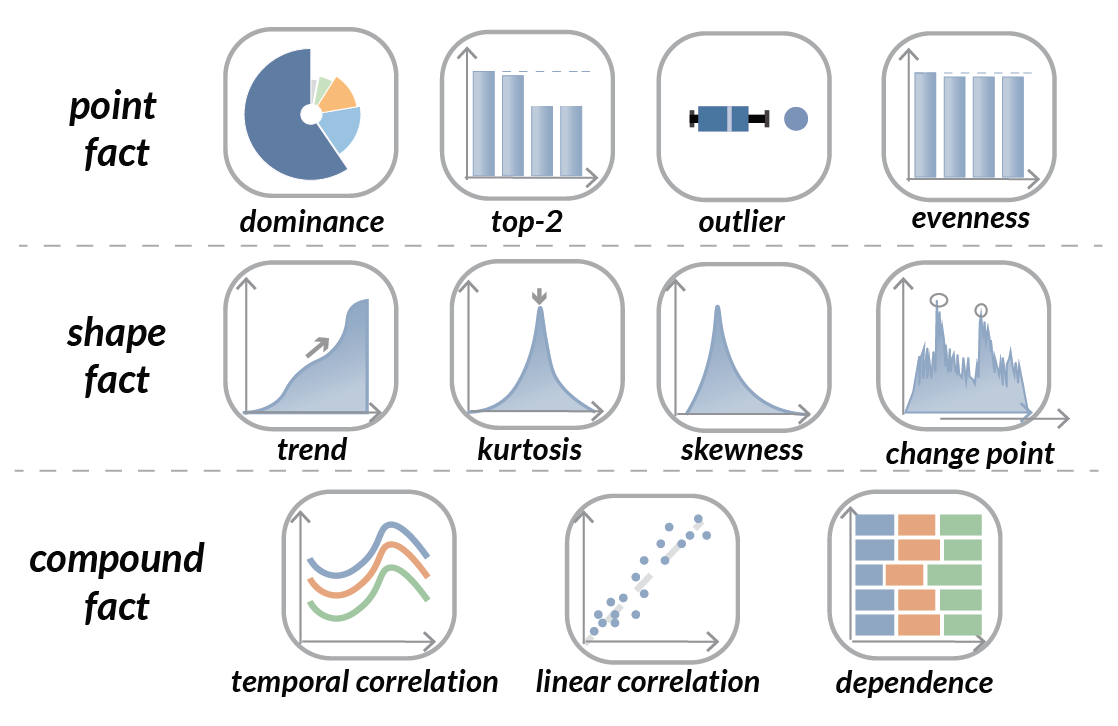}
    \caption{Three types of data facts: point type, shape type, and compound type.  
These are further categorized into eleven types.
}
    \label{fig:insights}
    \vspace{-1em}
\end{figure}

After performing data transformations, the original block data is converted into multiple data forms \( d \). Subsequently, we extract data facts from each data form \( d \) within the block. All extracted information is regarded as the data fact information contained within that block. 
By considering the characteristics of hierarchical tables and referencing definitions and classification criteria from existing studies~\citep{QuickInsights,CoInsight2024Li,TangBo2017}, we identified 11 different types of data facts to extract from \( d \). These data facts are grouped into three primary categories: point type, shape type, and compound type (see Fig.~\ref{fig:insights}). 
Based on these 11 data facts types, we sequentially analyze each data form within the block to determine the presence of these data facts. Subsequently, corresponding Vega-Lite~\citep{Vega-Lite} visualizations are generated and stored to intuitively present the data facts identified. Following this process, we obtain a collection of data facts within the block, accompanied by their corresponding visual representations.

\subsubsection{Layout Architecture Design} 
\label{sec:Insight_Layout}

Building on the previous data fact extraction method, we identify relevant data facts from blocks in hierarchical tables. 
Because blocks vary in size and position and each may correspond to multiple facts, the extraction process can yield a large number of facts. 
Managing such volume presents challenges in how to organize and present them effectively across different regions and categories.
This section therefore discusses layout methods.

Existing work falls into three categories: (1) listing data facts by user queries or recommendation results~\citep{DataShot,HaoIALP2024}, which struggle to clearly display large volumes of facts; (2) building a relationship network based on associations among data facts for holistic layout~\citep{CoInsight2024Li}, which lacks intuitive mapping to the corresponding regions in the table; and (3) embedding visual representations of data facts directly into tables to reduce cognitive load~\citep{InsigHTable}, which can cause overlap and harm readability when too many charts are embedded. Hence, dedicated layout techniques for embedding data facts in hierarchical tables are needed to support effective, large-scale organization and interaction.

\begin{figure}
    \centering
    \includegraphics[width=1.0\linewidth]{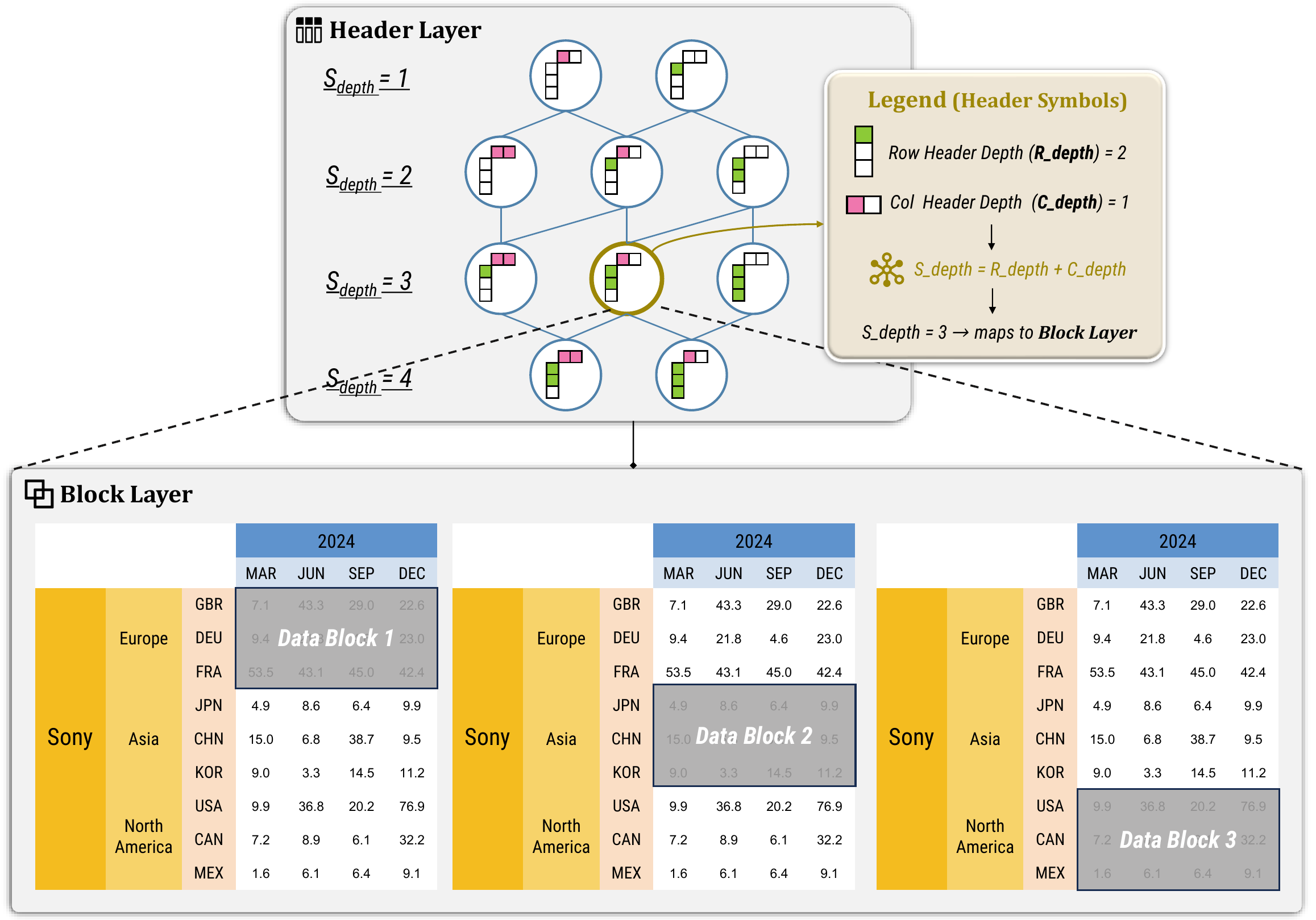}
    \caption{This figure illustrates the layout architecture of data facts. The architecture is divided into two layers. Header Layer describes the depth combination scheme of the row and column headers; Block Layer is used to determine the embedding position of each data block.
}
    \label{fig:3layer}
    \vspace{-1em}
\end{figure}

To address the complexity of embedding numerous data fact visualizations within hierarchical tables, we propose a data fact layout architecture, illustrated in Fig.~\ref{fig:3layer}. This architecture comprises, from top to bottom, the \textbf{Header Layer} and the \textbf{Block Layer}.

\textbf{Header Layer} is designed to systematically classify combinations of row and column header depths based on the hierarchical structure of the table headers in \( T_h \). For any given block within \( T_h \), its corresponding row and column header labels can be located within the hierarchical structures of row headers \( H_{\text{row}} \) and column headers \( H_{\text{col}} \). Typically, blocks at deeper header levels have smaller dimensions, containing fewer body cells. 
To quantify this relationship, we define \( S_{\text{depth}} \) as the sum of the row header depth \( R_{\text{depth}} \) and column header depth \( C_{\text{depth}} \) in a particular combination

Suppose that the row header structure \( H_{\text{row}} \) has a depth of \( r \), and the column header structure \( H_{\text{col}} \) has a depth of \( s \). Consequently, \( R_{\text{depth}} \) ranges from 0 to \( r \), and \( C_{\text{depth}} \) ranges from 0 to \( s \). 
In practice, multiple combinations of the depths of the headers of the row and column may yield the same \( S_{\text{depth}} \) value. For example, \( S_{\text{depth}} = 1 \) can correspond to \( (R_{\text{depth}} = 0, C_{\text{depth}} = 1) \) or \( (R_{\text{depth}} = 1, C_{\text{depth}} = 0) \). To visually differentiate these combinations, we propose a graphical representation, as depicted in the Header Layer in Fig.~\ref{fig:3layer}. Each circle represents a distinct depth combination scheme. 
The vertical squares on the left represent the row header levels of \( T_h \), while the horizontal squares on the top represent the column header levels. 
Shaded squares indicate the depths of the columns and rows in the current combination. 

Furthermore, previous research~\citep{CoInsight2024Li,Blocks_Eurovis_2022} introduced the concept of a parent-child relationship between blocks, representing inclusion relations among table blocks. 
Specifically, if one of the row or column headers of two blocks is identical while the other differs by exactly one level, with the deeper header structure identical to the shallower one except at the deepest level, then a parent-child relationship is established between these blocks. 
Using this concept, we incorporate a mapping relationship into the Header Layer. 
In detail, for a block corresponding to a particular \( S_{\text{depth}} = i \), the header depths of its parent block sum to \( S_{\text{depth}} = i - 1 \), with one of the depths (\( R_{\text{depth}} \) or \( C_{\text{depth}} \)) remaining constant. Consequently, in the Header Layer, we visually associate depth combinations exhibiting parent-child relationships using connecting lines between adjacent circles.
As shown in Fig.~\ref{fig:pipeline}(b), through structured classification and visual design, our method distinguishes the header depth combination schemes corresponding to different states, providing state navigation for users' data exploration.

\textbf{Block Layer} is primarily responsible for arranging various blocks based on specific combinations of row and column header depths, ensuring that each block is accurately placed within its corresponding region of the hierarchical table according to its row and column header labels.
Due to the inherent complexity of the header structures of rows and columns, hierarchical tables typically consist of numerous blocks. 
Directly positioning blocks with varying sizes and positions in the same table area inevitably results in overlaps and conflicts. 
Using the Header Layer classification approach, we ensure that only blocks that share identical \( R_{\text{depth}} \) and \( C_{\text{depth}} \) values are concurrently arranged within the same table region, mitigating overlaps and layout conflicts.

To precisely position each block within its designated table region, accurate coordinate information for each block must be determined. 
Using the hierarchical table parsing method described in Sec.~\ref{sec:Hierarchical-Table-Structure-Parsing}, each block's positional information, represented by row and column header labels, is obtained as \( B_{\text{loc}} = \{ R_{\text{loc}}, C_{\text{loc}} \} \). 
Consequently, we introduce a coordinate transformation function \( F_{\text{trans}} \), which receives the hierarchical structures of row and column headers \( H_{\text{row}} \) and \( H_{\text{col}} \), as well as the block position indicated by header labels \( R_{\text{loc}} \) and \( C_{\text{loc}} \), as inputs.
The coordinate transformation function \( F_{\text{trans}} \) calculates precise coordinates for blocks according to their corresponding row and column header labels, defined formally as:

\begin{equation}
F_{\text{trans}}(H_{\text{row}}, H_{\text{col}} ,R_{\text{loc}}, C_{\text{loc}}) = (x_1, x_2, y_1, y_2),
\end{equation}
here, \( (x_1, x_2) \) represents the range of rows derived from parsing the row header \( H_{\text{row}} \), and \( (y_1, y_2) \) denotes the range of columns obtained from parsing the column header \( H_{\text{col}} \). The resulting coordinate set \( (x_1, x_2, y_1, y_2) \) specifies the exact position of the block.

After the coordinate transformation described above, blocks with identical depth combinations are positioned into their corresponding table regions, ensuring an overlap-free layout. Next, we embed the relevant data fact visualizations into the corresponding block regions.
Using the data fact extraction method outlined in Sec.~\ref{sec:insight-extraction}, we analyze the transformed data within each block to identify relevant data facts from the predefined set of 11 types. For each identified data fact, we generate a visual chart representation. One issue arises during chart embedding: a single block region may correspond to multiple data facts, yet only one chart can be displayed at any given time, causing potential spatial conflicts. To address this, an alternative list is maintained to store charts that are not currently embedded.

\subsection{Applying Semantic Zooming}

Semantic Zooming is an interaction technique that dynamically adjusts the level of information displayed based on the zoom level, allowing users to view content with varying degrees of detail, effectively presenting hierarchical information structures to enhance user comprehension. 
In this section, we design a semantic zooming method tailored for the scenario of displaying data facts within hierarchical tables, enabling transitions between data facts at varying scales (Fig.~\ref{fig:pipeline}(c)).

Existing studies on semantic zooming have been applied in various fields, including large-scale data exploration~\citep{cubes,Big_Data_Landscapes}, geospatial visualization~\citep{COVID-19}, RDF~\citep{ZoomRDF}, and UML diagrams~\citep{UML}. \cqy{In table visualization, Slingsby et al. support multi-level exploration through zoom interactions~\citep{Aidan}:} each cell first presents a compact visualization of aggregated metrics, and semantic zooming progressively reveals details from overview to data points. However, this approach is designed for flat tables and cannot be directly adapted to hierarchical tables to dynamically switch and display data facts, highlighting the need for a customized semantic zooming method in hierarchical table contexts.

In Sec.~\ref{sec:Insight_Layout}, the proposed data fact layout architecture resolves layout conflicts during embedding. However, since different combinations of row–column header depths correspond to distinct views, interactive methods are still required to support view switching. 
To this end, we define a semantic zooming relationship based on \( S_{\text{depth}} \) and implement zoom-in and zoom-out interactions through the mouse wheel. 
During semantic zooming, the table size remains constant, while embedded charts dynamically adjust with the zoom level. 
A zoom-in indicates the user’s intention to inspect data facts at a finer granularity, while a zoom-out corresponds to a more macro-level perspective (Fig.~\ref{fig:pipeline}(c)). 
This visualization further reveals the hierarchical parent–child relationship between the original block and the zoomed-in block, allowing dynamic analysis of data facts at different scales within a block.

However, the view after semantic zooming is not uniquely determined. For example, consider the visualization of the data facts presented at \( S_{\text{depth}} = 2 \) (with \( R_{\text{depth}} = 1 \) and \( C_{\text{depth}} = 1 \)). 
Performing a zoom-in operation to \( S_{\text{depth}} = 3 \), two potential visualizations may emerge: one focusing on the increased depth of the row header (\( R_{\text{depth}} = 2 \), \( C_{\text{depth}} = 1 \)), and the other emphasizing the greater depth of the column header (\( R_{\text{depth}} = 1 \), \( C_{\text{depth}} = 2 \)). Furthermore, the state of view (that is, which data facts should be shown in each block) is not fixed and depends on the user’s current exploratory interests. Therefore, determining which state to display after a zoom-in operation is an important design consideration.

\begin{figure}
    \centering
    \includegraphics[width=0.9\linewidth]{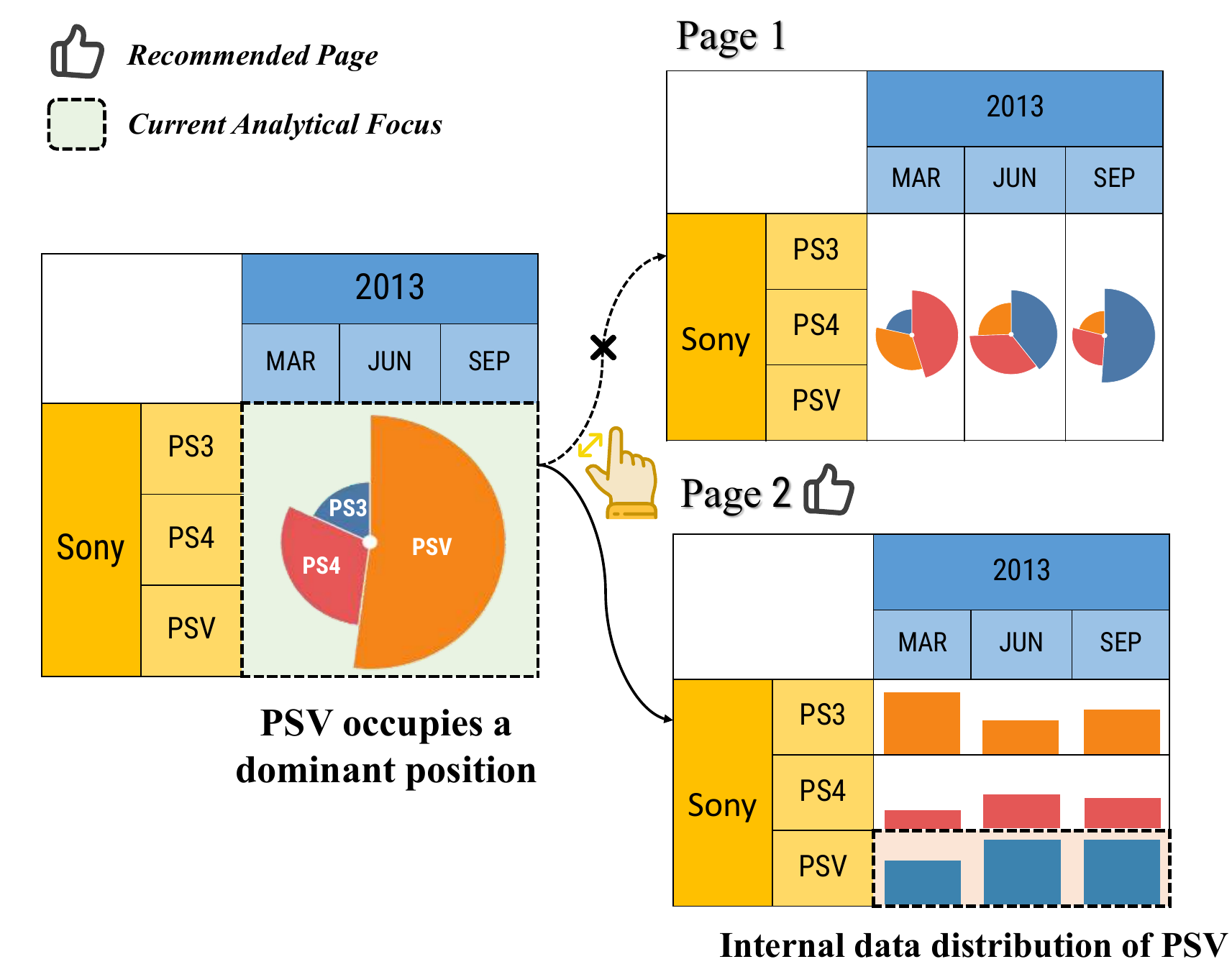}
    \caption{This figure illustrates the state recommendation process based on data fact features during semantic zooming interactions.
}
    \vspace{-1em}
    \label{fig:Recommend}
\end{figure}

To address this challenge, we propose a state recommendation strategy for inner data facts (see Fig.~\ref{fig:pipeline}(d)), where each distinct view corresponding to a specific combination of row and column header depths is termed a ``Page''. The goal of this strategy is to recommend and display the state that is most relevant to the data facts just explored after performing semantic zooming operations. 
For example, as illustrated in Fig.~\ref{fig:Recommend}, the left side depicts the initial state, where the currently focused block presents a pie chart illustrating the data fact ``dominance'', summarizing data aggregated from multiple rows to conclude that ``PSV occupies a dominant position''. When the user initiates a zoom-in operation, the system provides two alternatives: Page 1 and Page 2. Although the block on page 1 has a parent-child relationship with the initial block, page 1 emphasizes data distinctions along column headers, which are less relevant to the previously observed ``dominance'' data fact. In contrast, page 2 highlights the internal data distribution within the PSV row, closely aligned with the previously focused ``dominance''. Consequently, employing the proposed data fact-based state recommendation strategy prioritizes the display of Page 2 after zoom-in operation, enabling users to further explore intrinsic data distribution characteristics within PSV.

We provide a formal description of the proposed state recommendation strategy, which precisely defines the process of selecting the optimal state after zoom operations based on the current focused data fact. This formal description includes the following points:

\begin{itemize}
    \item \( P_{\text{cur}} \): the currently displayed page.
    \item \( F_{\text{sel}}(P_{\text{cur}}) \): features of the selected data fact in \( P_{\text{cur}} \).

    \item \( P = \{P_1, P_2, \dots, P_n\} \): the set of candidate pages available after a zoom-in or zoom-out operation.
    \item \( F_{\text{can}}(P_i) \): features of the data facts included in candidate page \( P_i \).
\end{itemize}

The objective of the state recommendation strategy is to recommend a suitable page 
\cqy{\( P_{\text{rec}} \in P \) to be displayed after performing the semantic zooming operation such that \( F_{\text{sel}}(P_{\text{rec}}) \) closely aligns with \( F_{\text{can}}(P_{\text{cur}}) \).} This can be formally expressed as:

\begin{equation}
P_{\text{rec}} = \arg \max_{P_i \in P} \left( \text{Sim}(F_{\text{sel}}(P_{\text{cur}}), F_{\text{can}}(P_i)) \right),
\end{equation}

\cqy{Here, the similarity function \( \text{Sim} \in [0,1]\) takes into account three factors:
(1) the consistency of data fact types, 
(2) the attributes of the data facts (e.g., the direction of data merging and the rows and columns where outliers are located), 
and (3) the textual description of the data facts, which is calculated based on cosine similarity.
}
Through this process, the page displayed after the semantic zooming operation can better align with the user’s current focus.
\begin{figure*}
    \centering
    \includegraphics[width=\linewidth]{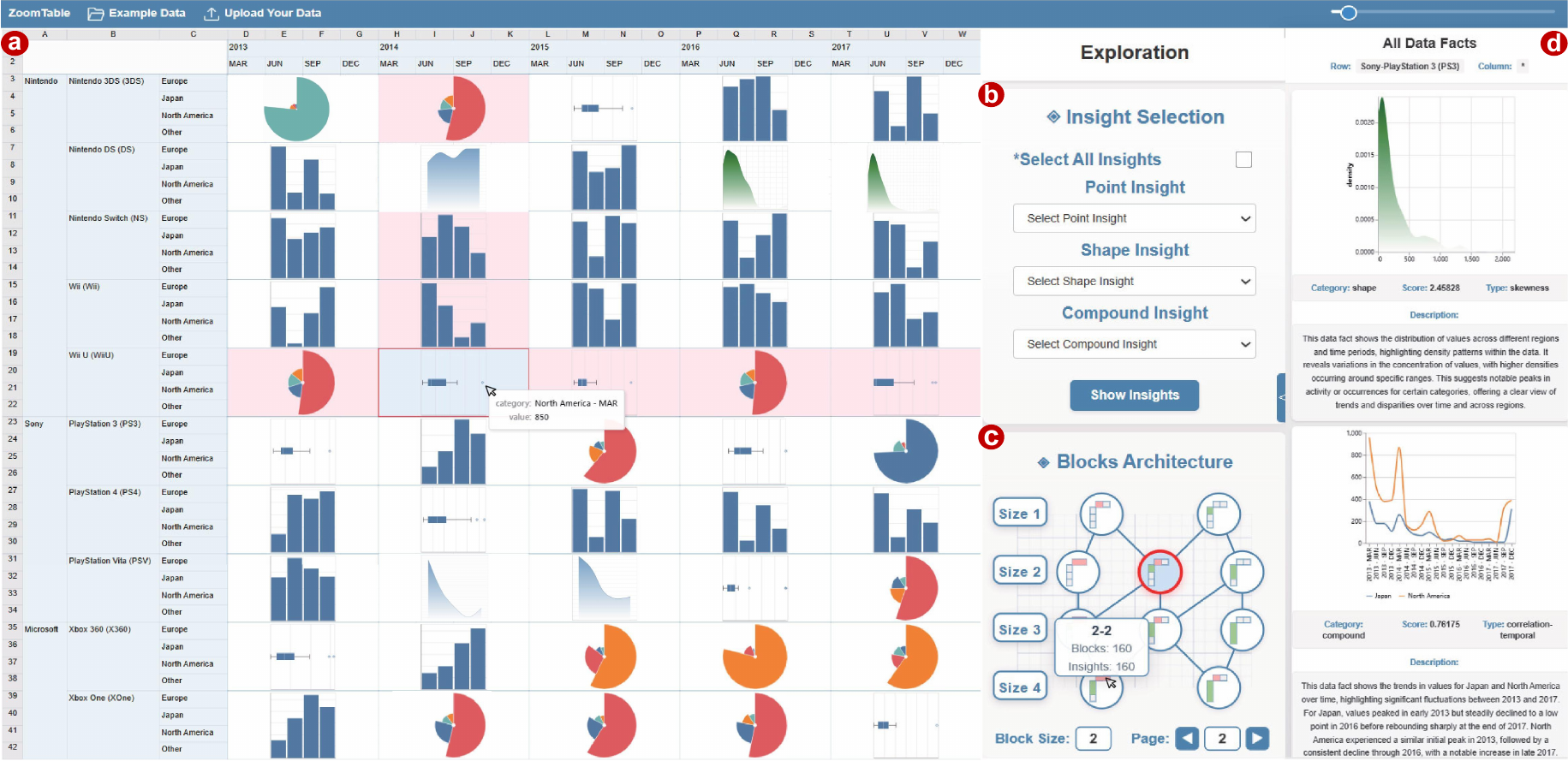}
    \caption{The ZoomTable user interface consists of four panels: (a) Hierarchical Table Interaction Panel: displaying the table and embedded charts, supporting exploration at multiple scales via semantic zooming. (b) Data Facts Filtering Panel: enabling personalized filtering of data fact types. (c) Navigation Panel: visualizing data fact layout and hierarchical relationships, assisting users in navigating across states. (d) Alternative Data Facts Panel: providing additional charts within the selected block for embedding and comparison.
}
    \label{fig:UI}
\end{figure*}

In addition to the automated state recommendation strategy, we implement a manual view-switching function, allowing users to transition between different views at the same \( S_{\text{depth}} \) level (see Fig.~\ref{fig:pipeline}(b)). Users can manually adjust the current page selection within the corresponding depth level. Based on the above functional design, users can, during the semantic zooming process, prioritize the analysis of recommended pages and manually adjust the pages and their data fact charts according to personalized analytical needs.

\section{The ZoomTable System}  
Based on the paradigm proposed in Sec.~\ref{sec:Method}, we further conduct unstructured interviews with several experts experienced in data analysis and visualization design to clarify the system's design considerations. Consequently, we develop the \system{} system, which enables the effective layout of data fact charts extracted from hierarchical tables and supports users in exploring them through semantic zooming.

\subsection{Design Considerations}
To define the functional objectives of the \system{} system, we conducted unstructured interviews with five domain experts and reviewed relevant studies to better clarify the design considerations. 
These experts include a university associate professor engaged in information visualization and human-computer interaction research (\textbf{E1}, age 33), two senior data scientists from a large Internet company (\textbf{E2}, age 35; \textbf{E3}, age 31), a visualization engineer working in a financial technology enterprise with a focus on tabular data analysis (\textbf{E4}, age 42), and a Ph.D. student whose research focuses on intelligent visualization recommendation (\textbf{E5}, age 28). 
All experts have more than three years of practical or research experience in the field of visual analytics or tabular data processing.
Their insights provide important references for us to refine the design considerations of the system, together with findings from existing studies, which ultimately lead to the identification of three \textbf{Design Considerations (DCs)} for the prototype system:

\textbf{DC1: Comprehensive Data Fact Extraction from Hierarchical Tables (E1, E3).}  
Hierarchical tables often organize information through multiple levels of headers in rows and columns. 
To support a thorough data analysis, the system must not only parse these hierarchical structures accurately but also extract all possible data facts embedded within them~\citep{Table_Illustrator_2024,Stolte_DC1_kdd02}. 
This involves identifying the hierarchical relationships among headers, aligning them with the corresponding data cells, and systematically generating data facts that preserve contextual meaning. 
This comprehensive extraction ensures that users can access the full analytical potential of the table.

\textbf{DC2: Intuitive Visual Design to Understand the Layout Architecture of Data Facts (E2, E4, E5).}  
The system should provide an intuitive and easy-to-understand visual design to help users clearly grasp the layout structure of data facts, thus optimizing the presentation of information and navigating the exploratory analysis processes. When dealing with complex data facts in hierarchical tables, the system must offer visualizations that reveal the relationships between the layout structure and the data facts~\citep{Design_Patterns_CHI18,Zheng_pvis2021_DC2,ChartStory_2023}. 

\textbf{DC3: Interactive Data Exploration through Semantic Zooming (E2, E3, E5).} 
The system should support flexible and interactive data exploration using semantic zooming. Specifically, it should adopt a layered information design that segments data into multiple levels of detail, allowing users to control the volume of information displayed and adjust the granularity of the visualization according to their analytical requirements~\citep{Cakmak_TVCG2022_DC3, Semantic_Zooming_Vitalis}. 

In summary, the three design considerations guide the development of the \system{} prototype system, aiming to enhance the understanding and exploration of data facts in hierarchical tables through precise parsing and presentation, intuitive visualization of data-fact layouts, and interactive exploration via semantic zooming.

\subsection{User Interface and Interaction}
This section outlines the user interface and interactions in the \system{} system, designed according to the previously outlined design considerations to ensure effective data exploration. Fig.~\ref{fig:UI} shows: (a) Hierarchical Table Interaction Panel, (b) Data Facts Filtering Panel, (c) Navigation Panel, and (d) Alternative Data Facts Panel. The specific functionalities and interaction designs are detailed below:

\system{} first parses uploaded hierarchical tables and renders the structure in panel (a) (Fig.~\ref{fig:UI}(a)).
To accomplish this, we have implemented two distinct table parsing modules in the system (\textbf{DC1}). 
The first traverses the hierarchy to infer block relations and extract data facts, independent of cell positions.
The second module reorganizes the original table into a structured data format conducive to rendering. 
Moreover, regarding embedded charts, the system ensures that the background dimensions match their areas of corresponding blocks, positioning each chart centrally within its block region. 
To enable rapid access to original data, right-click interactions are integrated into the chart area, facilitating rapid toggling between the charts and the underlying raw data. 
Recognizing potential readability issues with charts in smaller block areas, a global zooming feature is implemented, allowing users to uniformly adjust the scale of all elements within the table interface. 

The \system{} system incorporates a visualization and interaction strategy corresponding to the data facts layout architecture (\textbf{DC2}). The hierarchical structure illustrated from top to bottom in Fig.~\ref{fig:3layer} is mapped to panels (c) and (a) in Fig.~\ref{fig:UI}, respectively. 
Specifically, a visualization (see Fig.~\ref{fig:UI}(c)) represents combinations of row and column header depths within the hierarchical table. 
Here, ``block size'' denotes the sum of row and column header depths (\( S_{\text{depth}} \)), while ``page'' represents the specific view associated with a given block size, which supports manual view selection. 
Each circular node represents both a depth combination scheme and the corresponding view in the Hierarchical Table Interaction Panel. 
Once hierarchical information is acquired, these circular nodes visually display a grid glyph comprising green and red squares, denoting row and column header depths, respectively. 
Hovering over a node reveals details such as the number of data fact blocks and total data facts within the corresponding view.

Furthermore, the system records the view of the currently selected block, marking the corresponding node with a distinctive red border.
This red border remains visible even when switching views, provided that the block remains selected, thus aiding users in efficiently locating the relevant views.
Finally, we implement the display and switching functionality of the visual charts in the Alternative Data Facts Panel (see Fig.~\ref{fig:UI}(d)). 
Upon selection of a block in panel (a), the Alternative Data Facts Panel displays contextual information, including the block's position and all data fact charts within it, accompanied by detailed descriptions. 
Users can embed any selected chart into the main table area. Additionally, the Data Facts Filtering Panel (see Fig.~\ref{fig:UI}(b)) enables users to selectively filter displayed data fact types, such as exclusively showing ``Dominance'' type data facts. 

The Hierarchical Table Interaction Panel (see Fig.~\ref{fig:UI}(a)) serves as the primary data exploration interface, leveraging semantic zooming to facilitate cross-dimensional data fact exploration (\textbf{DC3}). 
Users also receive dynamic state recommendations aligned with the currently focused data fact.
Scrolling upwards over panel (a) initiates a zoom-in action, displaying finer-grained data facts, while scrolling downward triggers zoom-out, presenting broader data states. 
During semantic zooming, the visual indicators in the Navigation Panel (see Fig.~\ref{fig:UI}(c)) dynamically update to reflect the current state. 
Upon selecting a block region, the system extracts embedded data fact features and recommends an optimal adjacent view according to the recommendation strategy. 
Thus, subsequent semantic zoom actions directly transition users to recommended views, automatically updating relevant charts within new block regions. 
Additionally, modifying the embedded data fact visualization within a selected block triggers real-time updates to the recommended view, maintaining contextual consistency.

\section{Evaluation}
We evaluated the effectiveness of \system{} from two aspects based on hierarchical table data from the real world. 
On the one hand, we conduct a case study with a data analyst to examine the system’s application in real data fact exploration scenarios; on the other hand, we design user experiments to further validate the system’s performance and user experience through actual user interactions and feedback.

\subsection{Case Study} 
\label{sec:Case_Study}

\begin{figure*}
    \centering
    \includegraphics[width=1\linewidth]{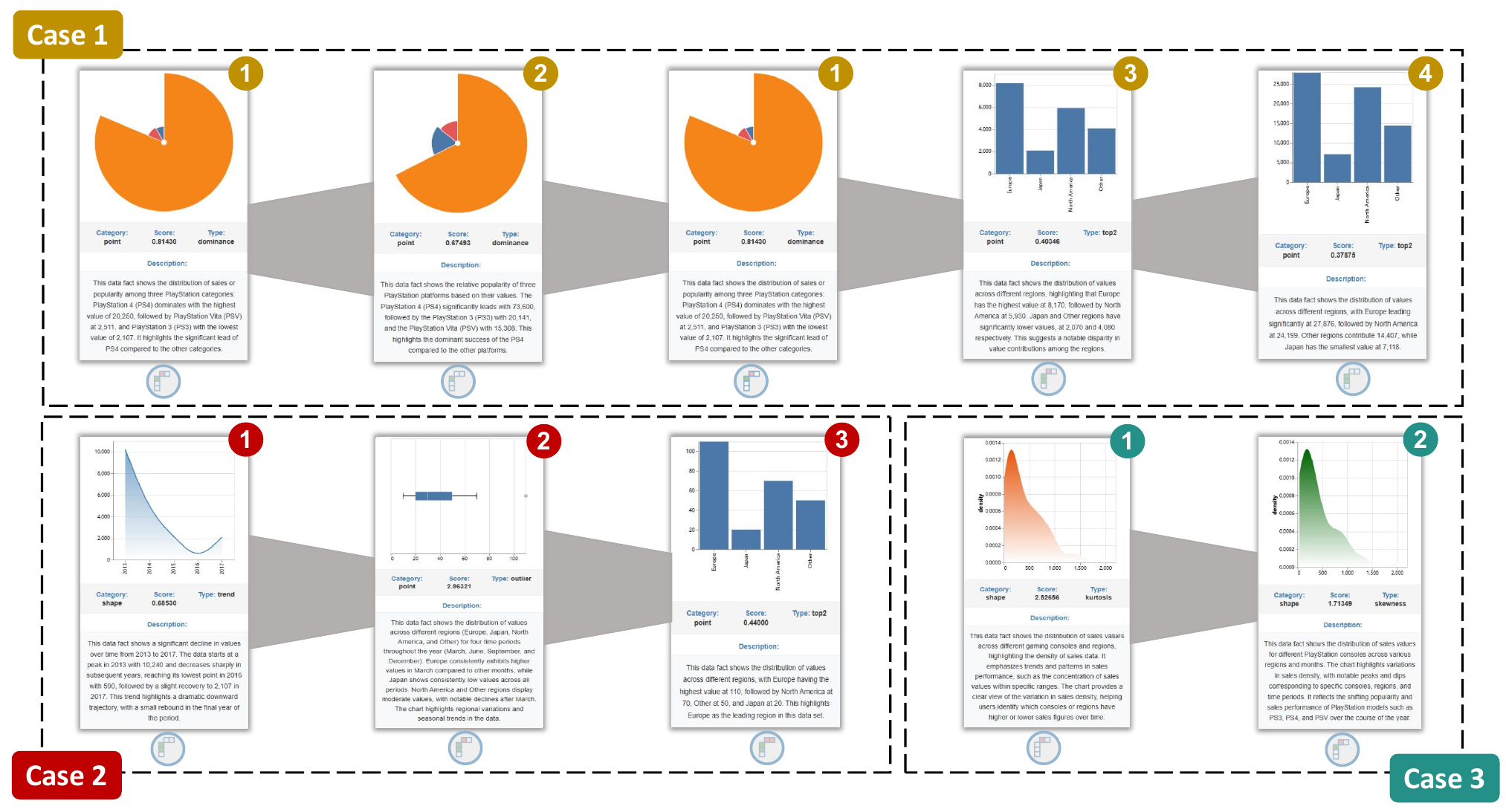}
    \caption{Case study example illustrating how the analyst incrementally explores data facts in a hierarchical sales dataset using ZoomTable. The figure presents three complete exploration paths, demonstrating the system’s effectiveness in real-world scenarios.}
    \label{fig:case}
\end{figure*}
We demonstrate the practical application of \system{} through a case study with an experienced data analyst (E4). The analyst works with a hierarchical dataset that records quarterly video game sales of multiple companies in countries and regions from 2013 to 2017, \cqy{in a table with 40 rows and 20 columns, featuring a three-level row header and a two-level column header.}

Before the study, we briefly introduce the main functionalities and visualization design of \system{}. 
The system parses the uploaded table. 
As data facts are not initially visualized, the analyst opts to display all facts (Fig.~\ref{fig:UI}(b)), after which the system updates the panels and presents the initial view. On this basis, the analyst reviews different views and selects three data facts of interest. 
Using semantic zooming, the analyst performs in-depth explorations, ultimately forming three exploration paths, as shown in Fig.~\ref{fig:case}.

\emph{Case 1} starts with a dominance-type data fact. The pie chart shows that ``PS4'' sales in 2017 were significantly higher than the other two consoles. 
To examine longer-term performance, the analyst zooms out. \system{} recommends and embeds a dominance chart of five-year total ``PS4'' sales, confirming sustained leadership.  
\cqy{Next, the analyst returns to the dominance data fact and zooms in to examine the internal factors underlying the 2017 dominance, drilling down into finer-grained segments.}
The system presents a top2-type data fact visualized using bar chart visualization, showing that in 2017, Europe and North America led in sales. 
To verify whether this pattern holds for a longer period of time, the analyst zooms out again. 
Based on the previously focused fact, \system{} recommends a same-type visualization for five-year regional totals, which indicates that Europe and North America consistently occupy the top two positions.

\emph{Case 2} starts with a trend-type data fact, shown as a line chart illustrating the sales trend of ``PS3'' from 2013 to 2017. The chart indicates that sales reached their lowest point in 2016. Therefore, the analyst performs a zoom-in interaction to further investigate. \system{} presents charts of ``PS3'' sales over years. 
The analyst notes that 2016 is an outlier: Although sales in Europe in the first quarter reached 110, overall sales were low, with a median of only 30. The analyst then performs another zoom-in interaction, focusing on first-quarter sales across regions. 
The results show that, despite overall low sales, Europe and North America still rank in the top two. 

\emph{Case 3} starts with a kurtosis-type data fact, which presents the sales distribution of different gaming consoles across regions in 2013. The analyst observes that the highest data density lies in the 0–500 range, while sales values above 1000 exhibit a very low density.
The analyst then zooms in. The 2013 table area is divided into three parts, each representing one company’s sales data. The analyst finds that Sony’s distribution is highly consistent with the data fact previously focused, with most sales still concentrated in the range of 0-500. Both data fact charts show a maximum density value of around 0.0013.

In summary, the case study demonstrates \system{}'s ability to explore complex hierarchical sales data. 
Using semantic zooming, the analyst uncovers sales trends and patterns, verifying the usability of the system in real-world scenarios.

\subsection{User Experiment}
To evaluate the performance of \system{} in data facts exploration, we design a user experiment consisting of a comparative experiment and an ablation study. First, we compare \system{} with vizGPT and CoInsight\citep{CoInsight2024Li} in terms of exploration efficiency and user experience: vizGPT generates data-fact charts through natural language instructions to support iterative exploration, while CoInsight automatically extracts data facts and constructs association graphs to enable exploration. We then conduct an ablation study. Since table parsing, data-fact presentation, and basic interaction are core components of the system and cannot be removed, we ablate three others instead: the recommendation strategy during semantic zooming (-w/o Rec), the filtering function in Fig.~\ref{fig:UI}(b) (-w/o Filt), and the layout visualization (Navigation Panel) in Fig.~\ref{fig:UI}(c) (-w/o Layout), to assess their impact on exploration efficiency.

\subsubsection{Experimental Setup}
We recruit 36 participants (12 female, 24 male), aged 21 to 44 years (M = 29.3), all of whom have regular experience in tabular data analysis. The sample consists of 18 industry practitioners (including data analysts, BI engineers, and finance/market researchers) and 18 university students (undergraduate and master’s level). Participants represent a diverse range of academic and professional backgrounds, including Computer Science, Automation, Economics, and Law. All participants report using spreadsheet or business intelligence tools for data analysis at least once per week, and none have prior exposure to \system{}, vizGPT, or CoInsight. To ensure the consistency of the experimental environment, all sessions take place in a controlled laboratory setting with identical devices and network configurations.

Based on a preliminary case study conducted by a data analyst, we confirm that the dataset used in Sec.~\ref{sec:Case_Study} contains rich data facts. Therefore, we adopt this dataset as the standardized dataset for the formal experiment and ensure that none of the participants has encountered it previously. In addition, we select two hierarchical tables from another dataset, HiTab~\citep{2022-hitab}, and provide them to the participants as training tests prior to the formal experimental tasks.

\subsubsection{Experimental Procedure}
In the comparative experiment, we consider that participants may develop prior knowledge of the dataset after using one tool, which could influence their performance when using subsequent tools. To address this, we employ a Latin square design to balance the order of system usage across groups. Specifically, the three systems used in the experiment are arranged into six different sequences following the Latin square design. A total of 36 participants are randomly assigned to these six sequences, with 6 participants per sequence. Each of the six experimental sessions is conducted independently to avoid cross-condition interference.
In the ablation study, participants are randomly and evenly divided into four groups, each corresponding to the original version of the system or one of the three ablated variants. This design enables us to compare the impact of removing specific functional modules on the efficiency of data fact exploration with the system.

\textbf{System Training}: Since none of the participants have previously been exposed to the two baseline systems or \system{}, we first provide user tutorials for the three systems to help participants understand their functionalities. We then explain key concepts to all participants, including data facts, table visualization, and semantic zooming. After the training, participants use the provided test dataset to explore the functionalities of the three systems and deepen their understanding of the relevant concepts. During this phase, we address any issues participants encounter with system usage or concept clarification. The entire system training and practice time is limited to 30 minutes.

\textbf{Task Assignment}: After completing the system training session, we assign the formal experimental tasks to the participants. First, the experimenter introduces the dataset used for the task, including its source and thematic content. Each group of participants is then asked to explore the dataset using their assigned system. The specific task is as follows: within the allotted time, participants need to locate content of interest from the table and use it as a starting point to iteratively expand, explore related data facts, and ultimately form coherent chains that are integrated into one or more complete data fact exploration paths, which are then recorded. In both the comparative experiment and the ablation study, each participant conducts the exploration with their assigned system or system variant, and the entire task session lasts 30 minutes. Participants are instructed to construct as many high-quality exploration paths as possible within the time limit, so as to comprehensively reflect their exploration ability with system support.

\textbf{Feedback Collection}: At this stage, participants are encouraged to actively discuss and raise questions, and we record all feedback in detail. To comprehensively evaluate the system’s user experience, we design a five-point Likert scale questionnaire covering two dimensions: the first assesses the system’s basic design, while the second focuses on participants’ experiences when performing specific data fact exploration tasks with the system.
In addition, we arrange two forms of interaction: first, a 15-minute group discussion conducted in a seminar-style setting, where participants can share their thoughts and provide feedback on their overall experience; second, a 5-minute individual interview with each participant to collect subjective experiences and personalized feedback based on their exploration process.

\subsubsection{Experimental Result}
\begin{figure}
    \centering
    \includegraphics[width=1\linewidth]{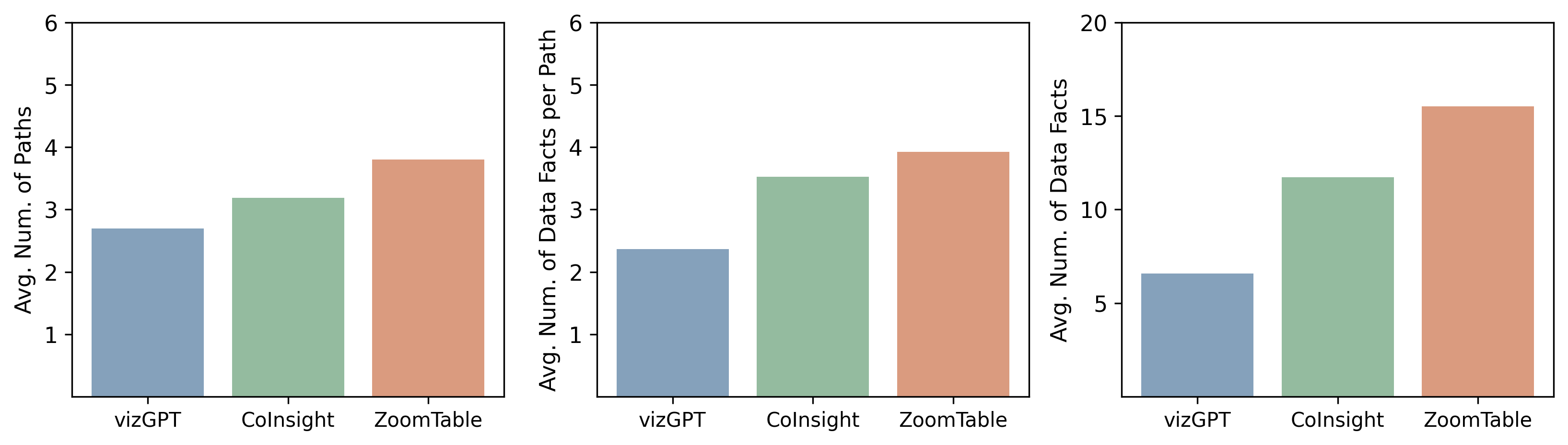}
    \caption{Quantitative performance of the three methods in the data facts exploration task. Left: Avg. Num. of exploration paths; Middle: Avg. Num. of data facts per path; Right: Avg. Num. of data facts explored.}
    \label{fig:result1}
    \vspace{-1em}
\end{figure}
In this section, we present the complete results of the user experiment, which consist of three parts: (1) quantitative statistics related to data fact exploration paths; (2) qualitative evaluation results based on a five-point Likert scale, including assessments of the system’s basic design as well as participants’ actual experiences when performing data fact exploration tasks; and (3) the results of the ablation study.

\textbf{Quantitative Evaluation.} Under identical time limits and a Latin-square balanced order, we compare the three methods with a focus on data-fact exploration. The quantitative metrics are the number of exploration paths per participant, the number of data facts per path, and the total number of data facts explored (see Fig.~\ref{fig:result1}). Results show that \system{} leads on all three: it yields more paths within the same time, more data facts per path, and a higher total number of discoveries. \cqy{Although CoInsight constructs a graph of data-fact associations, operating outside the original table context increases interpretation overhead and slows navigation, leading to fewer distinct exploration paths within the same time limit.} 

\begin{figure}
    \centering
    \includegraphics[width=1\linewidth]{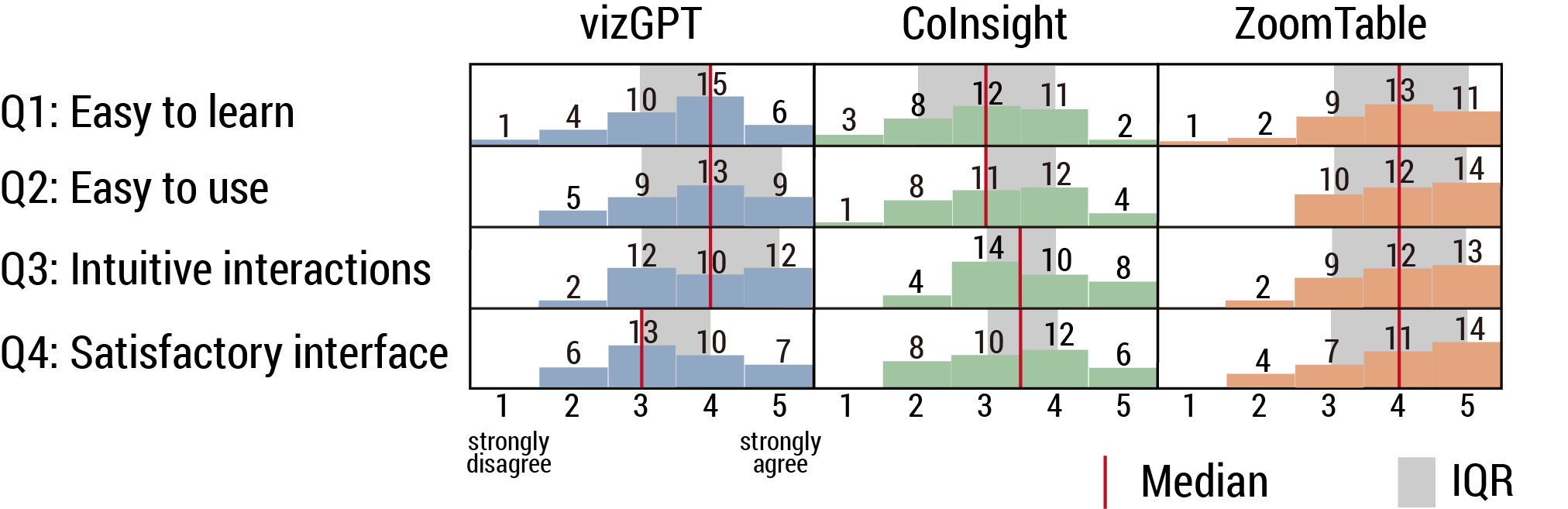}
    \caption{Participants’ ratings on basic system design across vizGPT, CoInsight, and \system{} (5-point Likert; medians and IQRs shown).}
    \label{fig:result2}
    \vspace{-1em}
\end{figure}

\begin{figure}
    \centering
    \includegraphics[width=1\linewidth]{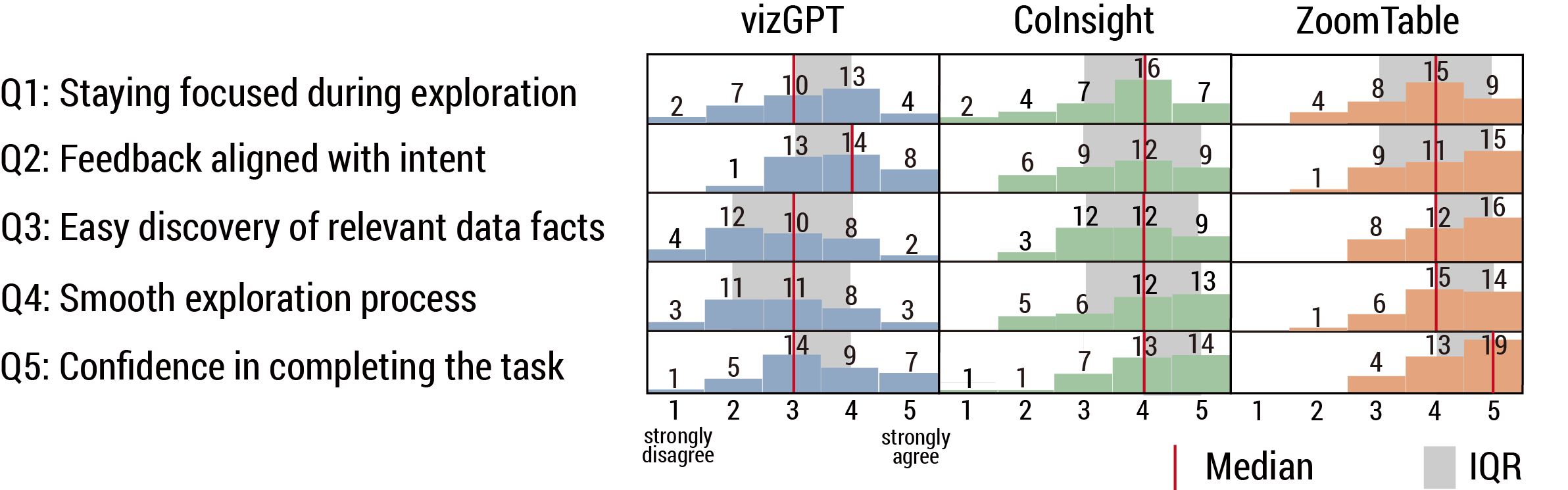}
    \caption{Participants’ ratings on the data-fact exploration experience across vizGPT, CoInsight, and \system{}.}
    \label{fig:result2_2}
    \vspace{-2em}
\end{figure}

\textbf{Qualitative Evaluation.} To ensure comparability, we use a unified two-dimensional 5-point Likert questionnaire. The first dimension assesses basic system design (see Fig.~\ref{fig:result2}); the second dimension assesses the experience of data-fact exploration tasks (see Fig.~\ref{fig:result2_2}). The results show that the three methods perform well in basic system design, with \system{} slightly ahead; its integrated table view layout reduces cognitive effort. For task experience, \system{} shows a clearer advantage: vizGPT relies on pure natural language turns, leading to interaction latency and higher uncertainty in intent alignment, thus scoring lower on \emph{smooth exploration} and \emph{confidence in completing the tasks}. Compared with CoInsight, \system{} embeds data facts within the original table and organizes multiscale views through semantic zooming. Its recommendation mechanism guides zoom targets, reducing view state switching, and preserving contextual continuity, leading to higher ratings on key items.

\begin{figure}
    \centering
    \includegraphics[width=1\linewidth]{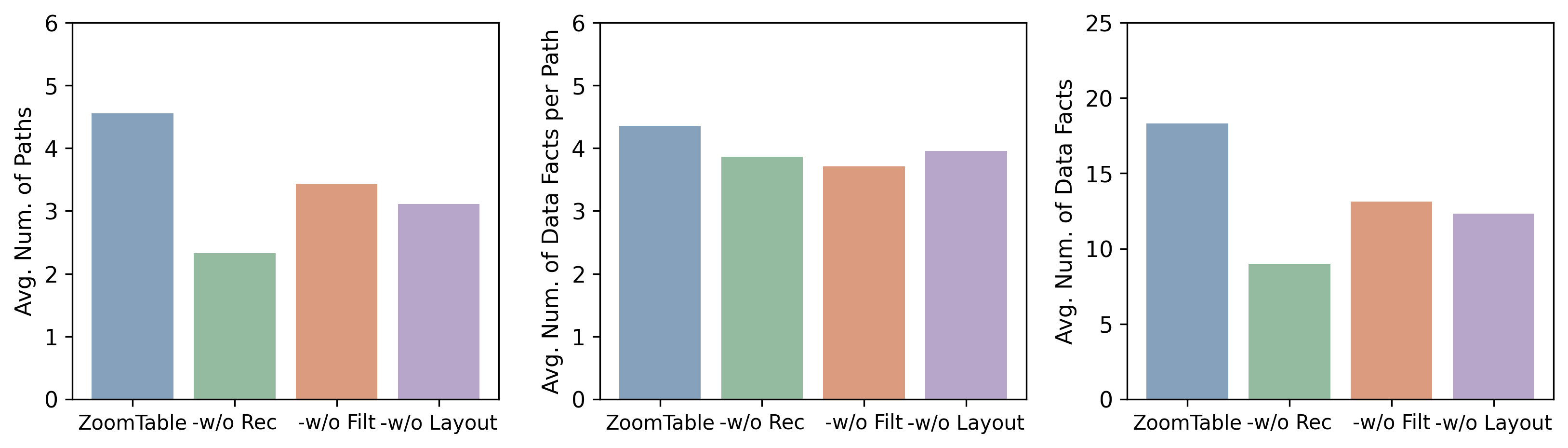}
    \caption{Ablation results for three variants: -w/o Rec, -w/o Filt, and -w/o Layout.}
    \label{fig:ablation_study}
    \vspace{-1em}
\end{figure}

\textbf{Ablation study.} To assess the contribution of key modules, we keep table parsing, data-fact presentation, and basic interactions unchanged, and remove in turn the state recommendation during semantic zooming (w/o Rec), data-fact type filtering (w/o Filt), and layout-structure visualization (w/o Layout). We compare each variant with the full system (Full) on the same three metrics as the main study (see Fig.~\ref{fig:ablation_study}). Results show that \textit{Full} achieves the highest values on all metrics (Avg. Num. of Paths, Avg. Num. of Data Facts per Path, and Avg. Num. of Data Facts). Participant feedback suggests the following mechanism: removing recommendation leads to the largest drop, indicating that the recommendation mechanism maintains alignment with the current focus and thus improves cross-scale exploration efficiency; removing layout-structure visualization weakens view localization and navigation, increasing cross-view back-and-forth; removing type filtering introduces more irrelevant candidates when focusing on a specific kind of data fact, reducing the effective yield per unit time.

\section{Discussion and Future Work}
Our study proposes a semantic zooming–based paradigm for exploring data facts in hierarchical tables and further develops the ZoomTable system. However, the system still presents several limitations in terms of interaction design, recommendation mechanisms, and research scope. Future work can proceed in the following three directions.

\textbf{Multi-Scale Semantic Zooming Optimization}.  
The current system leverages semantic zooming to present multiple layers of data facts. Through our analysis of existing methods, we identified that there is still room for improvement in the interactive semantic zooming process: first, enhancing the smoothness of semantic zooming and the responsiveness of interaction feedback to strengthen the continuity of cross-level exploration~\citep{Aidan}; and second, incorporating multi-scale semantic zooming to move beyond the single-scale approach that relies solely on header-based partitioning~\citep{cubes}. 

\textbf{Smart Recommendations \& Storytelling}.  
The current system recommends views during zooming based on the data facts users focus on, but there is still room for improvement. In the future, large language models could be used to better model contextual relationships among data facts, capture user intent more accurately, and enable more effective view transitions. In addition, the system could incorporate lightweight automated reporting and explanation functions, where exploration logs, embedded charts, and textual descriptions are automatically organized into structured outputs to help users generate more readable and narrative-rich reports~\citep{Storytelling_FutureWork_Zhao_2023,CodeToon}.

\textbf{Practical Significance of Hierarchical Tables}.  
Although this work focuses on hierarchical tables, they are not a niche data type but are widely used in practice, such as in business reports, educational statistics, and medical records\citep{2022-hitab}. Moreover, flat tables can be structurally converted into hierarchical ones, extending the applicability of the proposed method to broader tabular data scenarios.

\section{Conclusion}
We propose an interactive exploration paradigm for data facts in hierarchical tables based on semantic zooming and develop the ZoomTable prototype system. 
\cqy{The ZoomTable system parses hierarchical table structures and extracts multidimensional data facts, embedding them in situ within the original hierarchical table to preserve context, while the layered layout architecture manages layout conflicts when many facts are displayed at scale.}
The semantic zooming mechanism, combined with state recommendation, enables users to switch across hierarchical views and maintain the continuity of exploration paths. 
A case study verifies the applicability and usability of the ZoomTable system in real-world scenarios, and a user experiment further demonstrates that it outperforms existing methods in exploration efficiency.

\bibliographystyle{elsarticle-harv} %
\bibliography{refs}                 %

\end{document}